\documentclass{aastex}
\usepackage{graphicx}
\usepackage{times}
\begin{document}


\title{
BVRI Photometry of SN 2011fe in M101
}

\author{Michael W. Richmond}
\shorttitle{SN 2011fe in M101}
\shortauthors{M. W. Richmond and H. A. Smith }
\affil{
Physics Department, Rochester Institute of Technology,
84 Lomb Memorial Drive, Rochester, NY, 14623 USA
}
\email{mwrsps@rit.edu}
\author{Horace A. Smith}
\affil{
Department of Physics and Astronomy, Michigan State University, 
East Lansing, MI 48824, USA
}
\email{smith@pa.msu.edu}


\begin{abstract}

We present BVRI photometry of supernova 2011fe in M101
from $2.9$ to $182$ days after the explosion.
The light curves and color evolution show that 
SN 2011fe belongs to the ``normal'' subset of 
type Ia supernovae,
with 
$\Delta m_{15}(B) = 1.21 \pm 0.03$ mag.
After correcting for extinction
and adopting a distance modulus of 
$(m - M) = 29.10$ mag to M101,
we derive absolute magnitudes 
$M_B = -19.21$,
$M_V = -19.19$,
$M_R = -19.18$
and 
$M_I = -18.94$.
We compare visual
measurements of this event to our CCD photometry
and find evidence for a systematic difference
based on color.

%
\end{abstract}


\section{Introduction}
Supernova (SN) 2011fe in the galaxy M101 (NGC 5457) was discovered 
by the Palomar Transient Factory
(\citealt{Law2009}; \citealt{Rau2009})
in images taken on UT 2012 Aug 24
and announced later that day
\citep{Nuge2011a}.
As the closest and brightest
type Ia SN since SN 1972E
\citep{Kirs1973},
and moreover as one which appears to suffer
relatively little interstellar extinction,
this event should provide a wealth of 
information on the nature of thermonuclear
supernovae.

We present here photometry of SN 2011fe in the 
BVRI passbands obtained at two sites,
starting one day after the discovery and
continuing for a span of 179 days.
Section 2 describes our observational procedures,
our reduction of the raw images,
and the methods we used to extract instrumental magnitudes.
In section 3, we explain how the instrumental
quantities were transformed to the standard
Johnson-Cousins magnitude scale.
We illustrate the light curves and color curves
of SN 2011fe in section 4, 
comment briefly on their properties,
and discuss extinction along the line of sight.
In section 5, 
we examine the rich history of distance measurements
to M101
in order to choose a representative value with
which we then compute absolute magnitudes.
Using a very large set of visual measurements
from the AAVSO, 
we compare the visual and CCD V-band observations
in section 6.
We present our conclusions in section 7.

\section{Observations}

This paper contains measurements made at the RIT Observatory,
near Rochester, New York, 
and the Michigan State University (MSU) Campus Observatory,
near East Lansing, Michigan.
We will describe below the acquisition and reduction
of the images into instrumental magnitudes from each
site in turn.

The RIT Observatory is located on the campus of the 
Rochester Institute of Technology, at 
longitude 77:39:53 West, latitude +43:04:33 North,
and an elevation of 168 meters above sea level.
The city lights of Rochester make the northeastern 
sky especially bright, which at times affected our
measurements of SN 2011fe.
We used a Meade LX200 f/10 30-cm telescope
and SBIG ST-8E camera, which features a Kodak KAF1600 CCD chip
and astronomical filters made to the Bessell prescription;
with $3 \times 3$ binning,
the plate scale is 
$1{\rlap.}^{''}85$
per pixel.
To measure SN 2011fe, 
we took a series of 60-second unguided exposures through
each filter; the number of images
per filter ranged from 10, at early times, to 15 or 20 at late times.
We typically discarded a few images in each series
due to trailing.
We acquired dark and flatfield images each night,
switching from twilight sky flats to dome flats 
in late October.
The filter wheel often failed to return to its proper location
in the R-band, 
so, when necessary, we shifted the R-band flats
by a small amount in one dimension in order to match
the R-band target images.
We combined 10 dark images each night to create a master
dark frame, and 10 flatfield images in each filter to create
a master flatfield frame.
After applying the master dark and flatfield images in the
usual manner,
we examined each cleaned target image by eye.
We discarded trailed and blurry images and
measured the FWHM of those remaining.

\begin{figure}
 \plotone{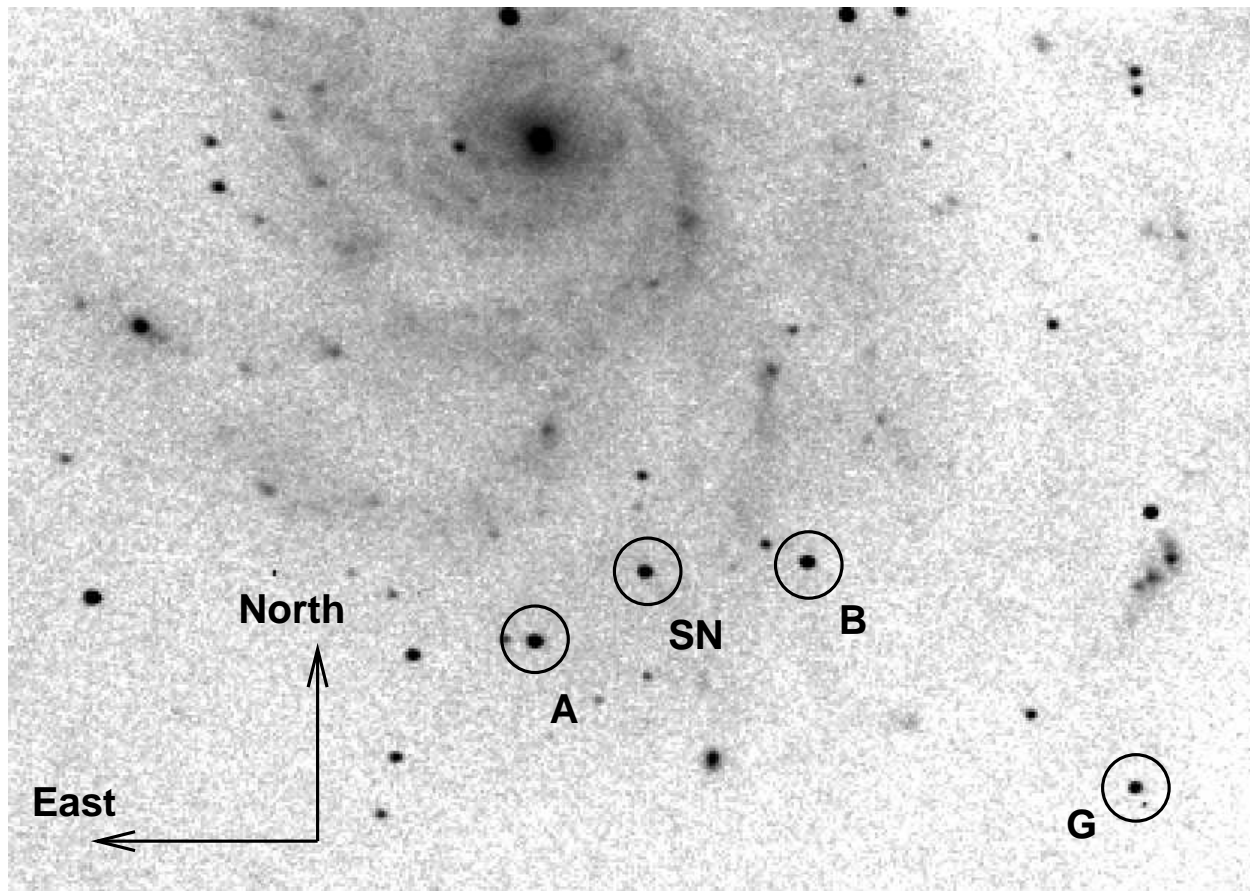}
 \caption{A V-band image of M101 from RIT,
          showing stars used to calibrate measurements 
          of SN 2011fe.  North is up, East to the left.
          The field of view is roughly 13 by 9 arcminutes.
          \label{fig:chartlabel} }
\end{figure}

The XVista 
\citep{Tref1989}
routines 
{\tt stars} 
and
{\tt phot}
were used to find stars 
and to extract their instrumental magnitudes,
respectively,
using a synthetic aperture with radius
slightly larger than the FWHM (which was typically 
$4''$ to $5''$).
As
Figure \ref{fig:chartlabel}
shows,
SN 2011fe lies in a region relatively free
of light from M101
(see also Supplementary Figure 1 of 
\citet{Li2011}).
As a check that 
simple aperture photometry would yield accurate results,
we examined high-resolution HST images of
the area,
using ACS WFC data in the F814W filter
originally taken as part of proposal GO-9490
(PI: Kuntz).
The brightest two sources within a $5''$ 
radius of the position of
the SN,
RA = 14:03:05.733,
Dec = +54:16:25.18 (J2000)
\citep{Li2011},
have apparent magnitudes 
of $m_I \simeq 21.8$ and 
$m_I \simeq 22.2$.
Thus, even when the SN is at its faintest,
in our final $I$-band measurements,
it is more than one hundred times brighter
than nearby stars which might contaminate our
measurements.

\begin{figure}
 \plotone{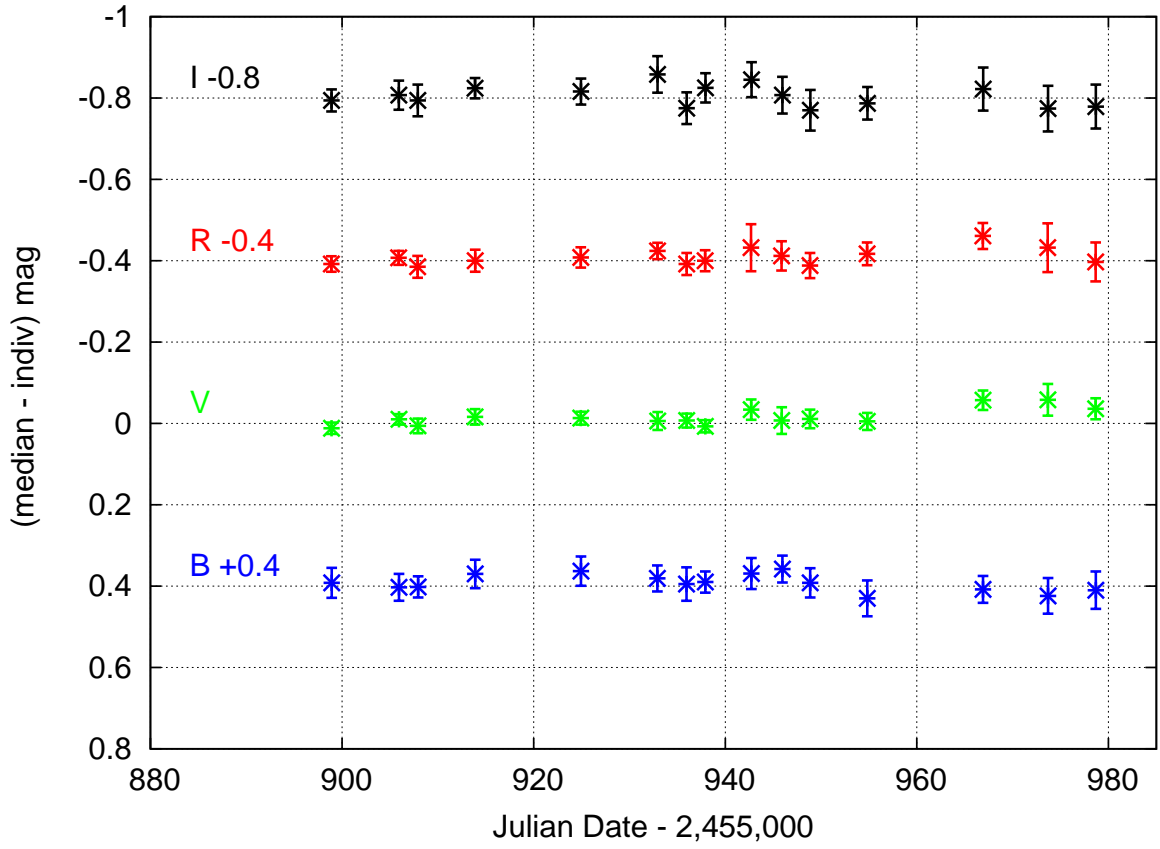}
 \caption{Difference between instrumental magnitudes
          extracted from median-combined images
          and from individual images at RIT. 
          The values have been shifted for clarity
          by 0.4, 0.0, -0.4, -0.8 mag in 
          B, V, R, I, respectively.
          \label{fig:indivmedian} }
\end{figure}

Between August and November, 2011, 
we measured instrumental magnitudes from each exposure
and applied inhomogeneous ensemble photometry
\citep{Hone1992}
to determine a mean value in each passband.
Starting in December, 2011,
the SN grew so faint in the I-band that
we combined the good images for each passband
using a pixel-by-pixel median procedure,
yielding a single image with lower noise levels.
We then extracted instrumental magnitudes
from this image in the manner described above.
In order to verify that this change in procedure
did not cause any systematic shift in the results,
we also measured magnitudes from the individual exposures,
reduced them using ensemble photometry,
and compared the results to those measured from the 
median-combined images.
As Figure \ref{fig:indivmedian} shows,
there were no significant systematic differences.

The Michigan State University Campus Observatory
lies on the MSU campus,
at longitude 05:37:56 West,
latitude +42:42:23 North,
and an elevation of 273 meters above sea level.
The f/8 60-cm Boller and Chivens reflector
focuses light on an Apogee Alta U47 camera
and its e2V CCD47-10 back-illuminated CCD,
yielding a plate scale of 0.56 arcseconds
per pixel.
Filters closely approximate the Bessell
prescription.
Exposure times ranged between 30 and 180 seconds.
We acquired dark, bias, and twilight sky flatfield
frames on most nights.
On a few nights, high clouds
prevented the taking of twilight sky flatfield exposures,
so we used flatfields from the preceding or following nights.
The I-band images show considerable fringing
which cannot always be removed perfectly.
We extracted instrumental magnitudes for all stars
using a synthetic aperture of radius 
$5{\rlap.}^{''}4$.

\section{Photometric calibration}

In order to transform our instrumental measurements
into magnitudes in the standard Johnson-Cousins BVRI
system,
we used a set of local comparison stars.
The AAVSO kindly supplied measurements for
stars in the field of M101 
\citep{Hend2012}
based on data from the 
K35 telescope at Sonoita Research Observatory
\citep{Simo2011}.
We list these magnitudes in Table
\ref{tab:compstars};
note that they are slightly different than the
values in the AAVSO's on-line sequences which
appeared in late 2011.
Figure \ref{fig:chartlabel}
shows the location of the three comparison stars.

\begin{center}
 \begin{table*}[ht]
   \caption{Photometry of comparison stars}
   \label{tab:compstars}
   {\small
    \hfill{}
    \begin{tabular}{l l l c c c c}
    \hline
    Star & RA (J2000) &  Dec (J2000) &  B  & V & R & I  \\
    \hline

    A  &  14:03:13.67 & +54:15:43.4 & $14.767 \pm 0.065$ & $13.832 \pm 0.027$ & $13.290 \pm 0.030$ & $12.725 \pm 0.037$ \\
    B  &  14:02:54.17 & +54:16:29.5 & $14.616 \pm 0.080$ & $13.986 \pm 0.037$ & $13.627 \pm 0.039$ & $13.262 \pm 0.044$ \\
    G  &  14:02:31.15 & +54:14:03.9 & $15.330 \pm 0.084$ & $14.642 \pm 0.042$ & $14.283 \pm 0.042$ & $13.931 \pm 0.073$ \\

    \hline
    \hline
  \end{tabular}
  }
 \end{table*}
\end{center}

The AAVSO calibration data included many other stars
in the region near M101.
In order to check for systematic errors,
we compared the AAVSO data to photoelectric BV measurements
in 
\citet{Sand1974}.
For the five stars listed as 
A, B, C, D and G in
\citet{Sand1974},
which range 
$12.01 < V < 16.22$,
we find mean differences of
$-0.013 \pm 0.038$ mag in B-band,
and
$-0.009 \pm 0.022$ mag in V-band.
We conclude that the AAVSO calibration set suffers
from no systematic error in B or V at the level of
two percent.
Unfortunately, we could not find any independent measurements
to check the R and I passbands in a similar manner.

In order to convert the RIT measurements to the Johnson-Cousins system,
we analyzed images of the standard field
PG1633+009 
\citep{Land1992}
to determine the coefficients in the transformation equations
\begin{eqnarray}
  B  &=  b  +  0.238 (043) * (b - v)  +  Z_B   \\
  V  &=  v  -  0.077 (010) * (v - r)  +  Z_V   \\
  R  &=  r  -  0.082 (038) * (r - i)  +  Z_R   \\
  I  &=  i  +  0.014 (013) * (r - i)  +  Z_I   
\end{eqnarray}
In the equations above, 
lower-case symbols represent instrumental magnitudes,
upper-case symbols Johnson-Cousins magnitudes,
terms in parentheses the uncertainties in each coefficient,
and $Z$ the zeropoint in each band.
We used stars A, B, and G to determine the 
zeropoint for each image
(except in a few cases for which G fell outside the image).
Table \ref{tab:ritphot} lists our calibrated measurements 
of SN 2011fe made at RIT.
The first column shows the mean Julian Date of all the exposures
taken during each night.
In most cases, the span between the first and last exposures was less than 
$0.04$ days, but on a few nights, clouds interrupted the sequence of 
observations.  
Contact the first author for a dataset providing the Julian Date 
of each measurement individually.

\begin{center}
 \begin{deluxetable}{l l l l l l}
   \tablecaption{RIT photometry of SN 2011fe \label{tab:ritphot}}
   \tablehead{
     \colhead{JD-2455000} &
     \colhead{B} &
     \colhead{V} &
     \colhead{R} &
     \colhead{I} &
     \colhead{comments}
   }

\startdata
    \\

   799.56  &  $14.072 \pm 0.038$  &  $13.776 \pm 0.016$  &  $13.728 \pm 0.011$  &  $13.696 \pm 0.022$  &   clouds \\
   800.58  &  $13.321 \pm 0.046$  &  $13.025 \pm 0.013$  &  $12.955 \pm 0.024$  &  $12.942 \pm 0.026$  & \\
   802.56  &  $12.148 \pm 0.028$  &  $12.049 \pm 0.011$  &  $11.941 \pm 0.020$  &  $11.882 \pm 0.022$  & \\
   803.55  &  $11.690 \pm 0.023$  &  $11.643 \pm 0.016$  &  $11.512 \pm 0.015$  &  $11.471 \pm 0.024$  & \\
   804.56  &  $11.310 \pm 0.025$  &  $11.300 \pm 0.009$  &  $11.170 \pm 0.012$  &  $11.147 \pm 0.017$  & \\
   806.54  &  \nodata                 &  $10.659 \pm 0.020$  &  $10.572 \pm 0.028$  &  $10.569 \pm 0.052$  &   clouds \\
   808.54  &  $10.346 \pm 0.045$  &  $10.466 \pm 0.029$  &  $10.336 \pm 0.011$  &  $10.402 \pm 0.025$  & \\
   814.53  &  $10.034 \pm 0.039$  &  $10.014 \pm 0.003$  &  $10.042 \pm 0.084$  &  $10.260 \pm 0.030$  &   clouds \\
   815.53  &  $ 9.981 \pm 0.015$  &  $10.012 \pm 0.012$  &  $10.011 \pm 0.025$  &  $10.320 \pm 0.027$  & \\
   816.53  &  $10.072 \pm 0.052$  &  $ 9.998 \pm 0.008$  &  $10.006 \pm 0.035$  &  $10.362 \pm 0.036$  & \\
   817.58  &  $10.060 \pm 0.072$  &  $ 9.903 \pm 0.059$  &  $10.031 \pm 0.031$  &  $10.307 \pm 0.019$  &   clouds \\
   820.53  &  $10.171 \pm 0.010$  &  $10.082 \pm 0.013$  &  $10.080 \pm 0.014$  &  $10.505 \pm 0.018$  &   clouds \\
   822.52  &  $10.326 \pm 0.017$  &  $10.134 \pm 0.006$  &  $10.181 \pm 0.002$  &  $10.630 \pm 0.026$  & \\
   823.53  &  $10.405 \pm 0.015$  &  $10.185 \pm 0.014$  &  $10.283 \pm 0.015$  &  $10.691 \pm 0.013$  & \\
   825.52  &  $10.623 \pm 0.030$  &  $10.311 \pm 0.009$  &  $10.428 \pm 0.027$  &  $10.840 \pm 0.018$  & \\
   827.51  &  $10.829 \pm 0.028$  &  $10.459 \pm 0.016$  &  $10.580 \pm 0.024$  &  $10.918 \pm 0.025$  & \\
   829.51  &  $11.043 \pm 0.057$  &  $10.574 \pm 0.019$  &  $10.655 \pm 0.017$  &  $10.898 \pm 0.020$  & \\
   830.52  &  $11.167 \pm 0.014$  &  $10.629 \pm 0.011$  &  $10.672 \pm 0.021$  &  $10.894 \pm 0.015$  & \\
   832.51  &  $11.423 \pm 0.058$  &  $10.739 \pm 0.011$  &  $10.731 \pm 0.012$  &  $10.855 \pm 0.018$  &   clouds \\
   839.53  &  $12.228 \pm 0.016$  &  $11.116 \pm 0.016$  &  $10.850 \pm 0.008$  &  $10.699 \pm 0.033$  &   clouds \\
   840.50  &  $12.312 \pm 0.039$  &  $11.180 \pm 0.013$  &  $10.865 \pm 0.024$  &  $10.661 \pm 0.026$  & \\
   841.50  &  $12.407 \pm 0.049$  &  $11.237 \pm 0.011$  &  $10.925 \pm 0.016$  &  $10.672 \pm 0.031$  & \\
   842.50  &  $12.503 \pm 0.038$  &  $11.294 \pm 0.006$  &  $10.958 \pm 0.015$  &  $10.680 \pm 0.026$  & \\
   844.50  &  $12.688 \pm 0.035$  &  $11.430 \pm 0.016$  &  $11.054 \pm 0.016$  &  $10.738 \pm 0.045$  &   clouds \\
   852.90  &  $13.098 \pm 0.021$  &  $11.921 \pm 0.014$  &  $11.615 \pm 0.033$  &  $11.255 \pm 0.022$  & \\
   858.48  &  $13.314 \pm 0.034$  &  $12.144 \pm 0.020$  &  $11.862 \pm 0.013$  &  $11.564 \pm 0.024$  &   clouds \\
   859.49  &  $13.290 \pm 0.016$  &  $12.157 \pm 0.010$  &  $11.879 \pm 0.012$  &  $11.634 \pm 0.047$  &   clouds \\
   864.90  &  $13.340 \pm 0.037$  &  $12.332 \pm 0.012$  &  $12.085 \pm 0.013$  &  $11.884 \pm 0.010$  & \\
   868.49  &  $13.364 \pm 0.017$  &  $12.441 \pm 0.014$  &  $12.190 \pm 0.016$  &  $12.038 \pm 0.040$  & \\
   870.90  &  $13.375 \pm 0.011$  &  $12.478 \pm 0.016$  &  $12.256 \pm 0.008$  &  $12.141 \pm 0.005$  & \\
   872.88  &  $13.371 \pm 0.041$  &  $12.509 \pm 0.016$  &  $12.343 \pm 0.021$  &  $12.179 \pm 0.008$  &   clouds \\
   883.92  &  $13.584 \pm 0.006$  &  $12.826 \pm 0.010$  &  $12.688 \pm 0.011$  &  $12.676 \pm 0.014$  & \\
   887.92  &  $13.598 \pm 0.012$  &  $12.927 \pm 0.006$  &  $12.796 \pm 0.005$  &  $12.832 \pm 0.013$  & \\
   889.92  &  $13.699 \pm 0.037$  &  $12.968 \pm 0.019$  &  $12.904 \pm 0.026$  &  $12.909 \pm 0.012$  & \\
   890.93  &  $13.658 \pm 0.021$  &  $12.988 \pm 0.019$  &  $12.902 \pm 0.011$  &  $12.957 \pm 0.018$  & \\
   898.91  &  $13.761 \pm 0.037$  &  $13.217 \pm 0.014$  &  $13.155 \pm 0.019$  &  $13.245 \pm 0.027$  &  \\
   905.94  &  $13.831 \pm 0.033$  &  $13.398 \pm 0.013$  &  $13.353 \pm 0.017$  &  $13.470 \pm 0.036$  &  \\
   907.92  &  $13.877 \pm 0.026$  &  $13.442 \pm 0.018$  &  $13.445 \pm 0.027$  &  $13.580 \pm 0.039$  &  \\
   913.89  &  $13.919 \pm 0.035$  &  $13.559 \pm 0.019$  &  $13.624 \pm 0.027$  &  $13.732 \pm 0.025$  &   clouds \\
   924.92  &  $14.078 \pm 0.036$  &  $13.811 \pm 0.016$  &  $13.945 \pm 0.025$  &  $14.066 \pm 0.032$  &   clouds \\
   932.94  &  $14.191 \pm 0.032$  &  $14.006 \pm 0.022$  &  $14.145 \pm 0.020$  &  $14.256 \pm 0.045$  &   clouds \\
   935.98  &  $14.219 \pm 0.041$  &  $14.054 \pm 0.017$  &  $14.280 \pm 0.027$  &  $14.384 \pm 0.039$  &   clouds \\
   937.92  &  $14.253 \pm 0.026$  &  $14.124 \pm 0.015$  &  $14.325 \pm 0.026$  &  $14.423 \pm 0.036$  &  \\
   942.72  &  $14.298 \pm 0.038$  &  $14.204 \pm 0.025$  &  $14.402 \pm 0.058$  &  $14.506 \pm 0.043$  &   clouds \\
   945.89  &  $14.377 \pm 0.033$  &  $14.272 \pm 0.033$  &  $14.523 \pm 0.036$  &  $14.627 \pm 0.045$  &   clouds \\
   948.86  &  $14.420 \pm 0.036$  &  $14.336 \pm 0.023$  &  $14.639 \pm 0.031$  &  $14.702 \pm 0.050$  &  \\
   954.82  &  $14.524 \pm 0.044$  &  $14.465 \pm 0.021$  &  $14.757 \pm 0.028$  &  $14.803 \pm 0.040$  &  \\
   966.90  &  $14.716 \pm 0.033$  &  $14.623 \pm 0.024$  &  $14.951 \pm 0.032$  &  $15.034 \pm 0.053$  &  \\
   973.67  &  $14.792 \pm 0.044$  &  $14.754 \pm 0.039$  &  $15.165 \pm 0.060$  &  $15.174 \pm 0.056$  &  \\
   978.67  &  $14.884 \pm 0.046$  &  $14.894 \pm 0.026$  &  $15.340 \pm 0.048$  &  $15.242 \pm 0.054$  &  \\

    \enddata
 \end{deluxetable}
\end{center}

The uncertainties listed in Table
\ref{tab:ritphot}
incorporate the uncertainties in instrumental magnitudes
and in the offset to shift the instrumental values
to the standard scale, added in quadrature.
As a check on their size,
we chose a region of the light curve,
$875 < {\rm JD} - 2455000 < 930$,
in which the magnitude appeared to be a linear
function of time.
We fit
a straight line to the measurements in each passband,
weighting each point based on its uncertainty;
the results are shown in 
Table \ref{tab:linear}.
The reduced $\chi^2$ values, between $0.9$
and $1.6$,
indicate that our uncertainties accurately reflect
the scatter from one night to the next.
The decline rate is smallest in the blue, but it is 
still, at roughly 130 days after explosion,
significantly faster than the $0.0098$ mag/day 
produced by the decay of ${\rm ^{56}Co}$.

\begin{center}
 \begin{deluxetable}{l c c }
   \tablecaption{Linear fit to light curves $2455875 < {\rm JD} < 2455930$  \label{tab:linear} }
   \tablehead{
     \colhead{Passband} &
     \colhead{slope (mag/day)} &
     \colhead{reduced $\chi^2$}
   }

   \startdata
   \\

    B  &  $0.0117 \pm 0.0006$  &   1.2 \\
    V  &  $0.0247 \pm 0.0004$  &   1.6 \\
    R  &  $0.0312 \pm 0.0004$  &   0.9 \\
    I  &  $0.0346 \pm 0.0006$  &   1.0 \\

    \enddata
 \end{deluxetable}
\end{center}

The MSU data were transformed in a similar way,
using only stars A and B.
The transformation equations for MSU were
\begin{eqnarray}
  B  &=  b  +  0.25 (0.03) * (b - v)  +  Z_B   \\
  V  &=  v  -  0.08 (0.02) * (b - v)  +  Z_V   \\
  I  &=  i  +  0.03 (0.02) * (v - i)  +  Z_I   
\end{eqnarray}
In the equations above, 
lower-case symbols represent instrumental magnitudes,
upper-case symbols Johnson-Cousins magnitudes,
terms in parentheses the uncertainties in each coefficient,
and $Z$ the zeropoint in each band.

Table \ref{tab:msuphot} lists our calibrated measurements 
of SN 2011fe made at MSU.
Due to the larger aperture of the MSU telescope,
exposure times were short enough that the range
between the first and last exposures on each night
was less than $0.01$ days.

\begin{center}
 \begin{deluxetable}{l l l l l l}
   \tablecaption{MSU photometry of SN 2011fe \label{tab:msuphot} }
   \tablehead{
     \colhead{JD-2455000} &
     \colhead{B} &
     \colhead{V} &
     \colhead{I} &
     \colhead{comments}
   }

\startdata
    \\

   801.58  &  $ 12.66 \pm  0.02$  &  $ 12.42 \pm  0.02$  &  $ 12.36 \pm  0.03$  &  \\  
   803.56  &  $ 11.69 \pm  0.01$  &  $ 11.60 \pm  0.01$  &  $ 11.48 \pm  0.02$  &  \\  
   806.58  &  $ 10.80 \pm  0.01$  &  $ 10.74 \pm  0.01$  &  $ 10.66 \pm  0.02$  &  \\  
   809.56  &  $ 10.36 \pm  0.02$  &  $ 10.30 \pm  0.01$  &  $ 10.28 \pm  0.02$  &  \\  
   811.56  &  $ 10.17 \pm  0.03$  &  $ 10.08 \pm  0.03$  &  $ 10.22 \pm  0.02$  &   clouds  \\  
   820.54  &  $ 10.13 \pm  0.02$  &  $ 10.06 \pm  0.01$  &  $ 10.44 \pm  0.02$  &  \\  
   822.53  &  $ 10.27 \pm  0.02$  &  $ 10.10 \pm  0.01$  &  $ 10.55 \pm  0.01$  &  \\  
   825.54  &  $ 10.55 \pm  0.03$  &  $ 10.27 \pm  0.02$  &  $ 10.87 \pm  0.02$  &   clouds  \\  
   837.52  &  $ 11.93 \pm  0.02$  &  $ 10.97 \pm  0.02$  &  $ 10.66 \pm  0.03$  &  \\  
   851.50  &  $ 13.04 \pm  0.03$  &  $ 11.83 \pm  0.02$  &  $ 11.16 \pm  0.03$  &  \\  
   857.90  &  $ 13.18 \pm  0.05$  &  $ 12.09 \pm  0.04$  &  $ 11.42 \pm  0.04$  &   clouds  \\  
   867.48  &  $ 13.27 \pm  0.04$  &  $ 12.37 \pm  0.03$  &  $ 11.96 \pm  0.03$  &  \\  
   889.46  &  $ 13.61 \pm  0.03$  &  $ 12.92 \pm  0.03$  &  $ 12.89 \pm  0.03$  &  \\  
   898.47  &  $ 13.67 \pm  0.10$  &  $ 13.14 \pm  0.06$  &  $ 13.22 \pm  0.08$  &  \\  

    \enddata
 \end{deluxetable}
\end{center}

\section{Light curves}

\begin{figure}
 \plotone{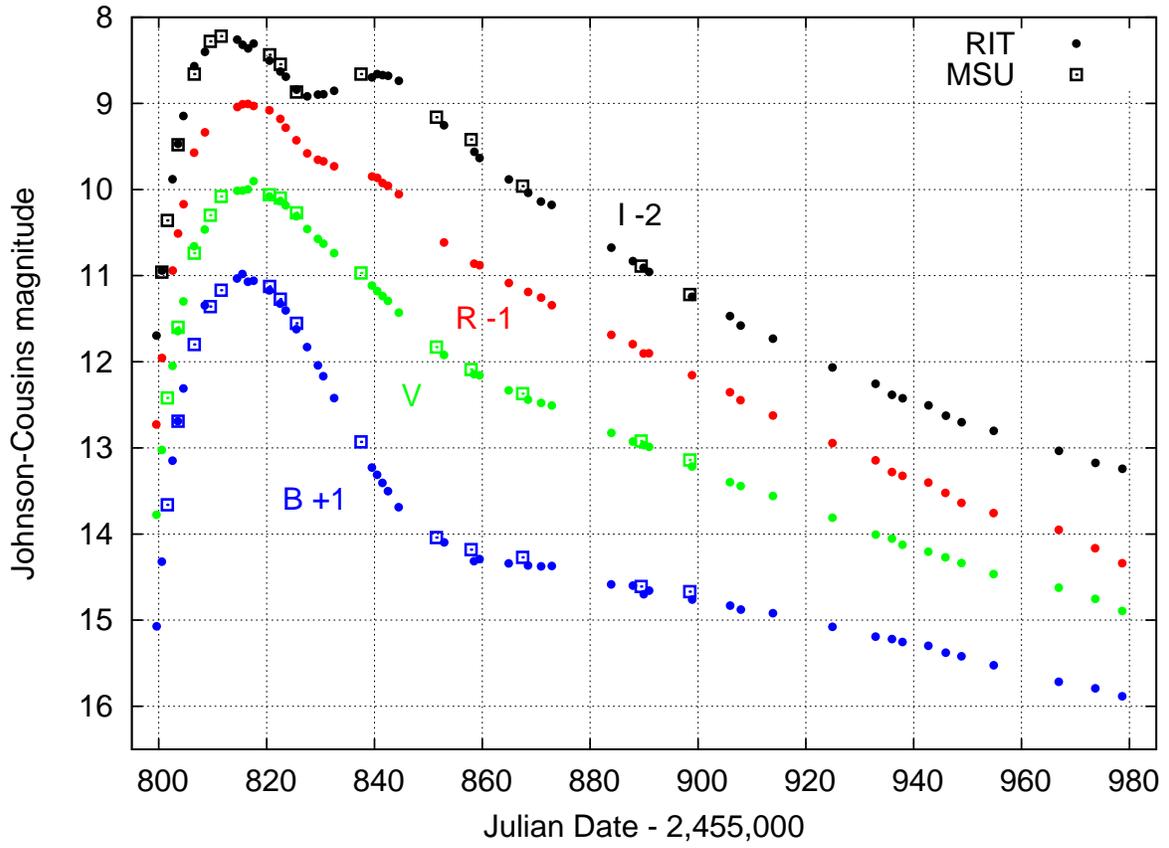}
 \caption{Light curves of SN 2011fe in BVRI.
          The data for each passband have
          been offset vertically for clarity.
          \label{fig:bvricurves} }
\end{figure}

We adopt the explosion date of
${\rm JD} \ 2455796.687 \pm 0.014$ 
deduced by 
\citet{Nuge2011b}
in the following discussion.
Figure \ref{fig:bvricurves}
shows our light curves of SN 2011fe,
which start 
$2.9$ days after the explosion
and 
$1.1$ days after 
\citet{Nuge2011a}
announced its discovery.

In order to determine the time and magnitude at 
peak brightness,
we fit polynomials of order 2 and 3 
to the light curves near maximum in each passband,
weighting the fits by the uncertainties in each measurement.
We list the results in 
Table \ref{tab:appmax},
including the values for the secondary maximum in $I$-band.
We again use low-order polynomial fits to measure
the decline in the $B$-band 15 days after the peak,
finding
${\Delta}_{15}(B) = 1.21 \pm 0.03$.
This value is similar to that of the ``normal''
SNe Ia 
1980N
\citep{Hamu1991},
1989B
\citep{Well1994},
1994D
\citep{Rich1995}
and 2003du
\citep{Stan2007}.
The 
location of the secondary
peak in $I$-band,
$26.6 \pm 0.5$ days after
and $0.45 \pm 0.03$ mag below the primary peak,
also lies close to the values for those other SNe.

Although there is as yet little published analysis
of the spectra of SN 2011fe,
\citet{Nuge2011b} 
state the the optical
spectrum on UT 2011 Aug 25 resembles that
of the SN 1994D; on the other hand,
\citet{Mari2011}
reports that a near-infrared spectrum on
UT 2011 Aug 26 resembles that of SNe Ia
with fast decline rates and 
$\Delta m_{15}(B) > 1.3$.
We must wait for detailed analysis of 
spectra of this event as it evolves
to and past maximum light for a 
secure spectral classification,
but this very preliminary information
may support the photometric evidence
that SN 2011fe falls into the 
normal subset of type Ia SNe.

\begin{center}
 \begin{deluxetable}{l c r }
   \tablecaption{Apparent magnitudes at maximum light \label{tab:appmax} }
   \tablehead{
     \colhead{Passband} &
     \colhead{JD-2455000} &
     \colhead{mag}
   }

   \startdata
   \\

    B        &  $816.0  \pm 0.3$  &   $10.00 \pm 0.02$ \\
    V        &  $817.0  \pm 0.3$  &   $ 9.99 \pm 0.01$ \\
    R        &  $816.6  \pm 0.4$  &   $ 9.99 \pm 0.02$ \\
    I        &  $813.1  \pm 0.4$  &   $10.21 \pm 0.03$ \\
    I (sec)  &  $839.7  \pm 0.5$  &   $10.66 \pm 0.01$ \\

    \enddata
 \end{deluxetable}
\end{center}

We turn now to the evolution of SN 2011fe in color.
In order to compare its colors easily to those of other supernovae,
we must remove the effects of extinction due to gas and dust
within the Milky Way and within M101.
Fortunately, there appears to be little intervening
material.
\citet{Schl1998}
use infrared maps of dust in the Milky Way to estimate
$E(B-V) = 0.009$
in the direction of M101.
\citet{Pata2011} 
acquired high-resolution spectroscopy of SN 2011fe
and identified a number of narrow 
Na I ${\rm D_2}$ 
absorption features;
they use radial velocities to assign some to the Milky Way
and some to M101.
They convert the
total equivalent width of all components,
$85 {\rm m\AA}$,
to a reddening of
$E(B-V) = 0.025 \pm 0.003$
using the relationship given in 
\citet{Muna1997}.
Note, however, that this total equivalent width 
is considerably smaller 
than that of all but a single star in the sample used
by 
\citet{Muna1997},
so we have decided to double the quoted uncertainty.
Adopting the conversions from reddening to extinction
given in 
\citet{Schl1998},
we compute the extinction toward SN 2011fe to be
$A_B = 0.11 \pm 0.03$,
$A_V = 0.08 \pm 0.02$,
$A_R = 0.07 \pm 0.02$,
and
$A_I = 0.05 \pm 0.01$.

\begin{figure}
 \plotone{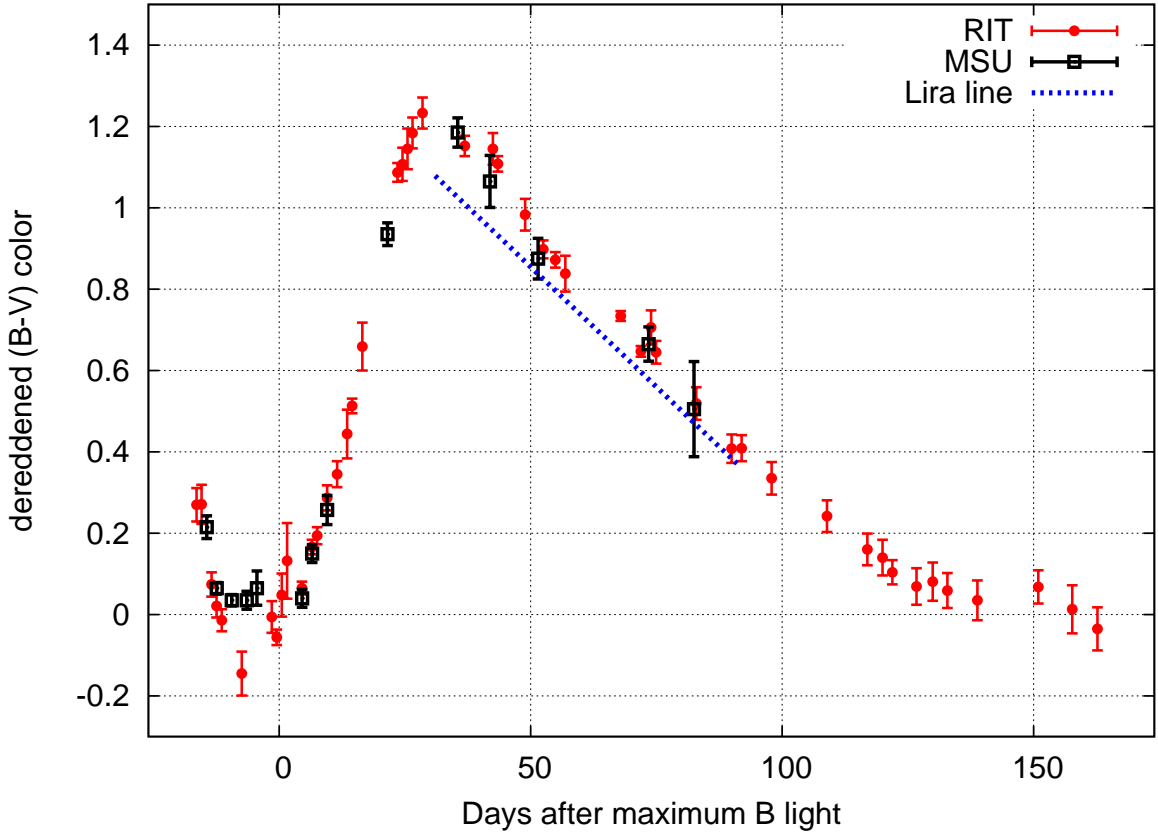}
 \caption{(B-V) color evolution of SN 2011fe,
          after correcting for extinction.
          \label{fig:bvcolor} }
\end{figure}

\begin{figure}
 \plotone{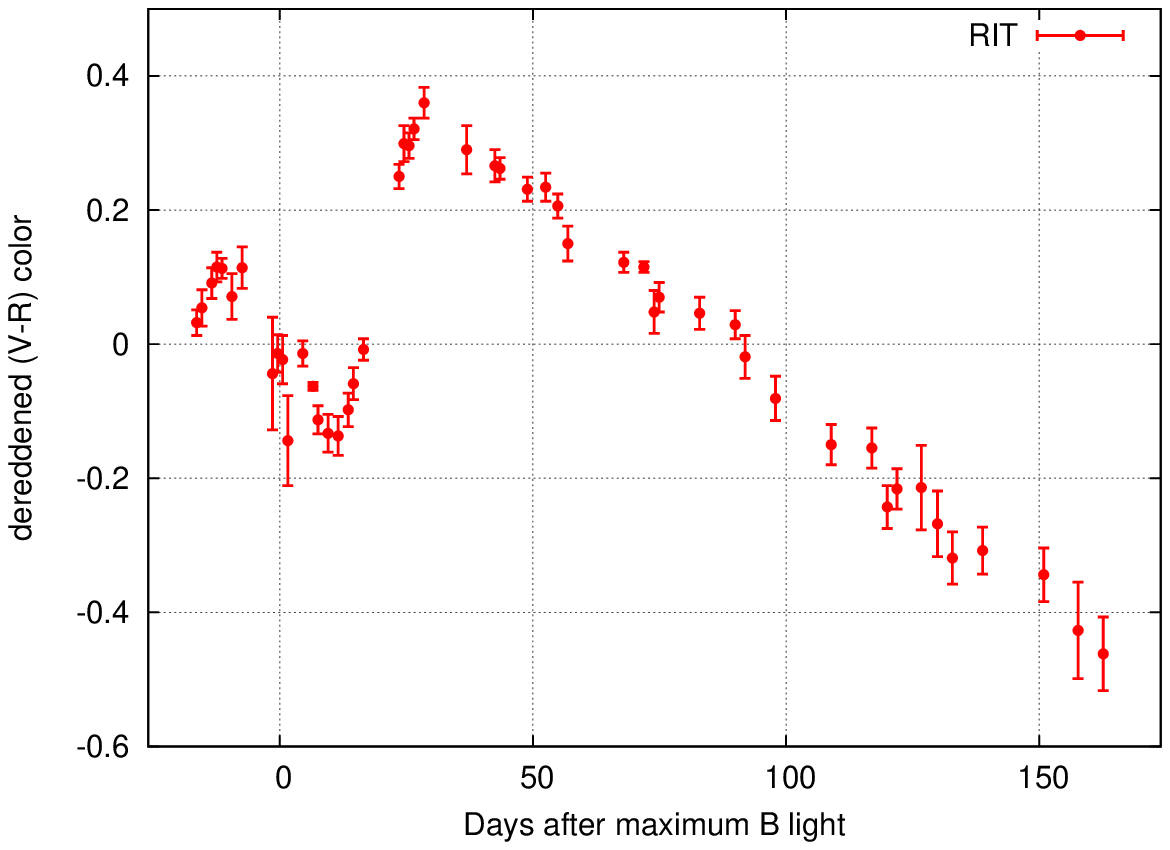}
 \caption{(V-R) color evolution of SN 2011fe,
          after correcting for extinction.
          \label{fig:ritvrcolor} }
\end{figure}

\begin{figure}
 \plotone{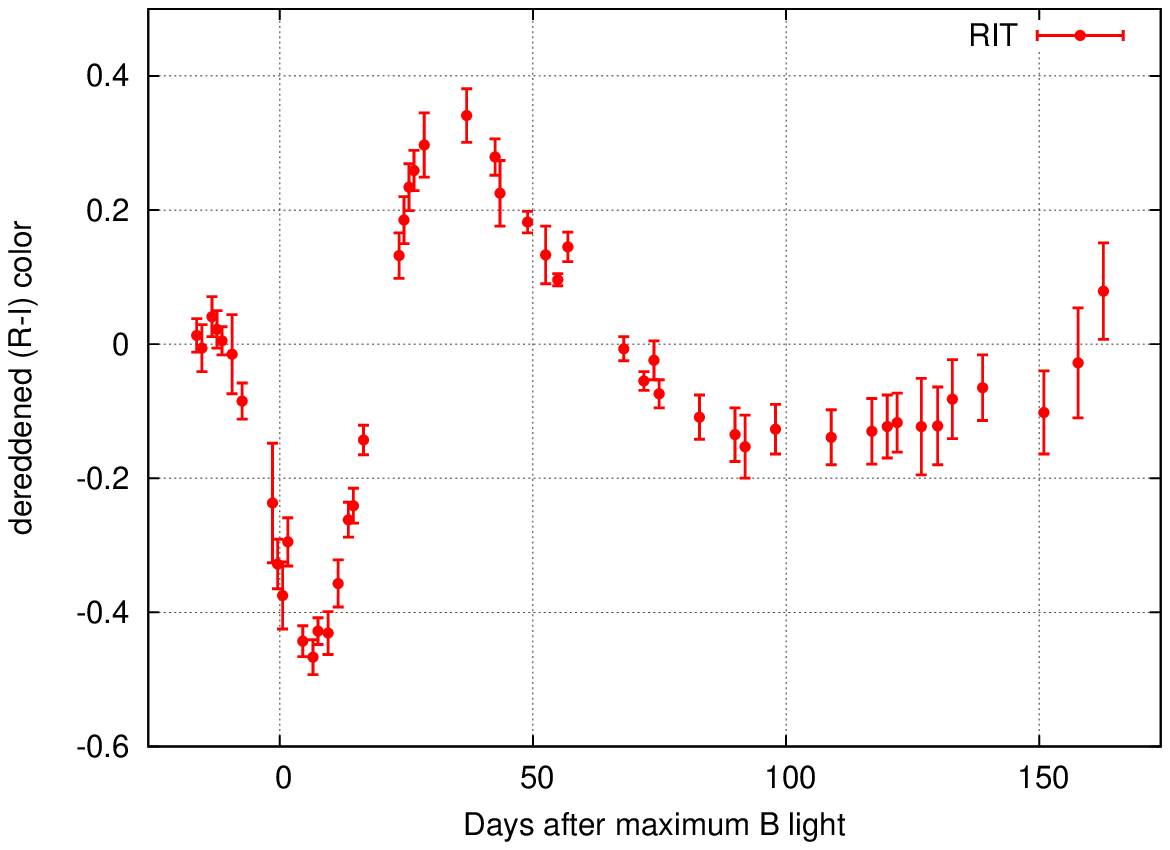}
 \caption{(R-I) color evolution of SN 2011fe,
          after correcting for extinction.
          \label{fig:ritricolor} }
\end{figure}

After removing this extinction from our measurements,
we show the color evolution of SN 2011fe
in Figures 
\ref{fig:bvcolor} ---
\ref{fig:ritricolor}.
The shape and extreme values of these colors
are similar to those of the normal Type Ia
SNe 1994D and 2003du.
In Figure
\ref{fig:bvcolor},
we have drawn a line to represent the relationship
(\citet{Lira1995} ; \citet{Phil1999})
for a set of four type Ia SNe which suffered
little or no extinction.
The $(B-V)$ locus of SN 2011fe lies slightly 
(0.05 to 0.10 mag)
to the red side of this line,
especially near the time of maximum $(B-V)$ color.
Given our estimates of the extinction 
to SN 2011fe, 
this small difference is unlikely to be due
to our underestimation of the reddening.

\section{Absolute magnitudes}

In order to compute the peak absolute magnitudes of SN 2011fe,
we must remove the effects of extinction and
apply the appropriate correction for its distance.
The previous section discusses the extinction
to this event,
and we now examine the distance to M101.
Since the first identification of Cepheids in this
galaxy 26 years ago
\citep{Cook1986},
astronomers have acquired ever deeper and larger
collections of measurements.
\citet{Shap2011}
provide a list of recent efforts,
which suggests that Cepheid-based measurements
are converging on a relative distance modulus
$(m - M) = 10.63$ mag between the LMC and M101.
If we adopt a distance modulus of
$(m - M)_{\rm LMC} = 18.50$ mag to the LMC,
this implies a distance modulus $(m - M)_{\rm M101} = 29.13$
to M101.
This is similar to one 
of the two 
results based on the TRGB method,
$(m - M)_{\rm M101} = 29.05 \pm 0.06 ({\rm rand}) \pm 0.12 ({\rm sys})$ mag
\citep{Shap2011},
though considerably less than the 
other,
$(m - M)_{\rm M101} = 29.42 \pm 0.11$ mag
\citep{Saka2004}.
We therefore adopt
a value of
$(m - M)_{\rm M101} = 29.10 \pm 0.15$ mag
to convert our apparent to absolute magnitudes.
Note that the uncertainty in this distance modulus
is our rough average,
based on a combination of the random and systematic
errors quoted by other authors and the scatter
between their values.
This uncertainty in the distance to M101
will dominate the uncertainties in all
absolute magnitudes computed below.

Using this distance modulus, 
and the extinction derived earlier for each band,
we can convert the apparent magnitudes at maximum light
into absolute magnitudes.
We list these values in 
Table \ref{tab:absmax}.

\begin{center}
 \begin{deluxetable}{l r r}
   \tablecaption{Absolute magnitudes at maximum light, corrected for extinction \label{tab:absmax} }
   \tablehead{
     \colhead{Passband} &
     \colhead{observed mag\tablenotemark{a} } &
     \colhead{based on $\Delta m_{15}$ \tablenotemark{b} }
   }

   \startdata
   \\

    B         &   $-19.21 \pm 0.15$  &  $-19.25 \pm 0.03$ \\
    V         &   $-19.19 \pm 0.15$  &  $-19.18 \pm 0.03$ \\
    R         &   $-19.18 \pm 0.15$  &  $-19.19 \pm 0.04$ \\
    I         &   $-18.94 \pm 0.15$  &  $-18.92 \pm 0.03$ \\
    I (sec)   &   $-18.49 \pm 0.15$  &  \nodata \\

    \enddata
    \tablenotetext{a}{based on $(m - M)_{\rm M101} = 29.10 \pm 0.15 $ mag}
    \tablenotetext{b}{using the relationship from \citet{Prie2006} }
 \end{deluxetable}
\end{center}

\citet{Phil1993}
found a connection between the absolute magnitude 
of a type Ia SN and the rate at which it declines
after maximum:
quickly-declining events are intrinsically less luminous.
Further investigation 
(\citet{Hamu1996} ; \citet{Ries1996} ; \citet{Perl1997})
confirmed this relationship
and spawned several different methods to quantify it.
We adopt the 
$\Delta m_{15}(B)$ method, 
which characterizes an event by the 
change in its B-band luminosity 
in the 15 days after from maximum light.
The light curve of SN 2011fe yields
$\Delta m_{15}(B) = 1.21 \pm 0.03$ mag,
placing it in the middle of 
the range of values for SNe Ia.
\citet{Prie2006} 
compute linear relationships between the
$\Delta m_{15}(B)$ 
and peak absolute magnitudes
for a large sample of SNe.
If we insert our value of 
$\Delta m_{15}(B)$ 
into the equations from their
Table 3 for host galaxies with small reddening,
we derive the absolute magnitudes
shown in the rightmost column of 
Table \ref{tab:absmax}.
The excellent agreement with the observed 
values suggests that our choice of
distance modulus to M101 
may be a good one.

\section{Comparison with visual measurements }

Perhaps because it was the brightest SN Ia to appear
in the sky since 1972,
SN 2011fe was observed intensively by many astronomers.
The AAVSO received over 900 visual measurements 
of the event within six months of the explosion.
Since it was observed so well with both human eyes
and CCDs, 
this star provides an ideal opportunity to compare
the two detectors quantitatively.

\begin{figure}
 \plotone{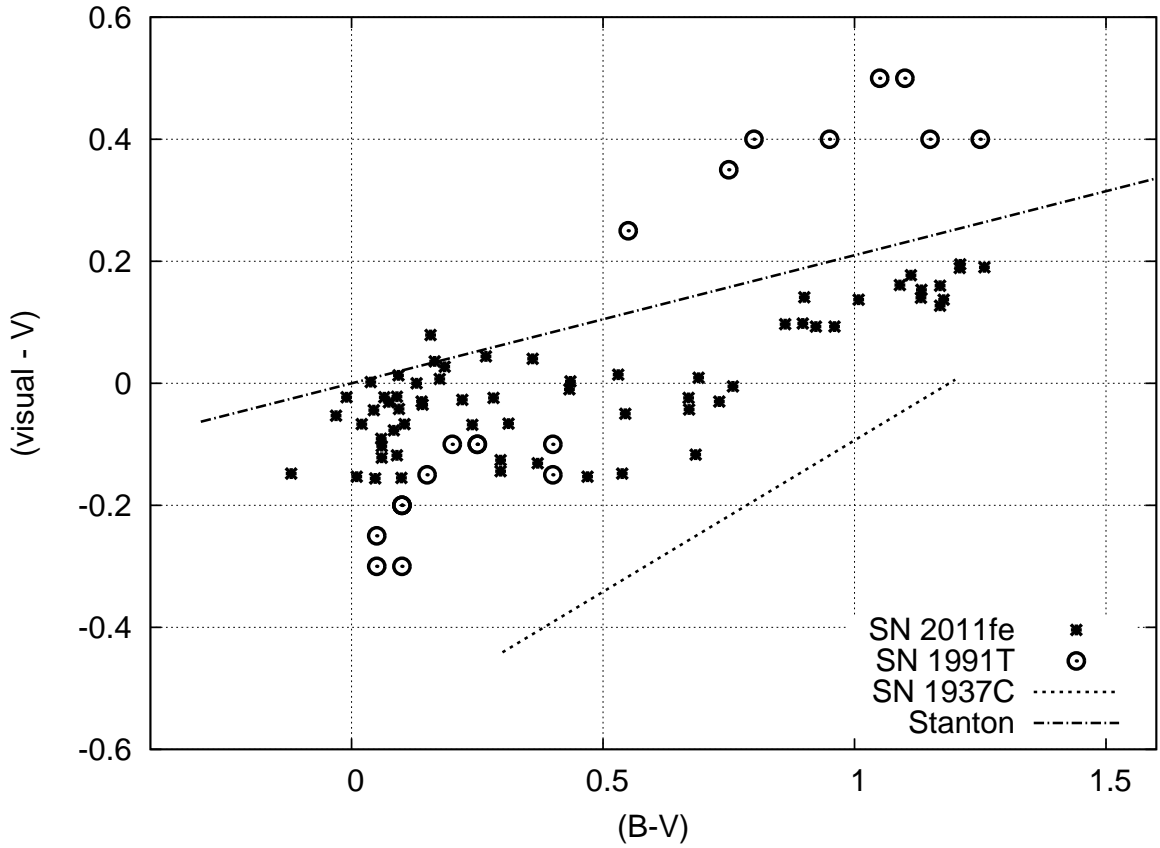}
 \caption{The difference between visual and 
          CCD or photographic measurements,
          as a function of (B-V) color,
          for SNe Ia and for variable stars in general.
          \label{fig:comparevisual} }
\end{figure}

We acquired visual measurements made by a large set of observers
from the AAVSO;
note that these have not yet been validated.
We removed a small number of obvious outliers,
leaving 880 measurements over the range
$799 < {\rm JD} - 2455000 < 984$.
For each of our CCD V-band measurements,
we estimated a simultaneous visual magnitude 
by fitting an unweighted low-order polynomial to the visual
measurements within $N$ days;
due to the decreasing frequency of visual measurements
and the less sharply changing light curve at late times,
we increased $N$ from 5 days to 8 days 
at JD $2455840$ and again to 30 days at
JD $2455865$.
We then computed the difference between the polynomial
and the V-band measurement.
Figure 
\ref{fig:comparevisual}
shows our results:
there is a clear trend for the visual measurements 
to be relatively fainter when the 
object is red.
If we make an unweighted linear fit to all the differences,
we find
\begin{equation}
  ({\rm visual - V})_{\rm 2011fe}  =  -0.09 + 0.19 (04) * (B-V) \\
\end{equation}
where the number in parentheses represents the uncertainty
in the coefficient.

We know of two other cases in which visual and 
other measurements of type Ia SNe are compared.
\citet{Pier1995} 
retrieved photographic films of SN 1937C,
which were originally described in 
\citet{Baad1938},
re-measured them with a photodensitometer,
and calibrated the results to the 
Johnson $V$-band using a set of local standards.
They compared their results to the visual
measurements of SN 1937C made by 
\citet{Beye1939} and found
\begin{equation}
 ({\rm visual - V})_{\rm 1937C} = -0.63 + 0.53 * (B-V) \\
\end{equation}
We plot this relationship in 
Figure
\ref{fig:comparevisual}
using a dotted line.
\citet{Jaco1996}
discussed the differences between 
visual measurements of SN 1991T from the AAVSO
to CCD $V$-band measurements made by
\citet{Phil1992}.
We have extracted the measurements of
\citet{Phil1992}
from their Figure 2 
and compared them to the visual measurements,
using the median of all visual measurements
within a range of 0.5 days to define a value
corresponding to each CCD measurement.
We show these differences as circular symbols
in 
Figure
\ref{fig:comparevisual};
an unweighted linear fit yields
\begin{equation}
 ({\rm visual - V})_{\rm 1991T} = -0.28 + 0.68(10) * (B-V) \\
\end{equation}

We find the slope to be the more interesting
quantity in these relationships, since the constant
offset term may depend on the choice of
comparison stars for visual observers.
Although at first blush the slopes appear to be quite
different, 
if one examines 
Figure
\ref{fig:comparevisual}
carefully,
one will see that the trend is quite similar
for all three SNe if one restricts the color
range to $(B-V) > 0.5$.
The main difference between these three events,
then, lies in the measurements made when the
SNe were relatively blue.
Could that difference be real?
We note that SNe 1991T (definitely) and 1937C (probably)
were events with slowly declining
light curves and higher than average luminosities,
while SN 2011fe declined at an average rate
and, for our assumed distance to M101,
was of average luminosity.
As 
\citet{Phil1992}
describes,
the spectrum of SN 1991T was most different
from that of ordinary SNe Ia at early times,
before and during its maximum luminosity;
it is also at these early times that SNe
shine with 1 light.
Could the combination of photometry by the
human eye and photometry by CCD
really distinguish ordinary and superluminous
SNe Ia at early times?
The evidence is far too weak at this time to
support such a conclusion, 
but we look forward to testing the idea
with future events.

\citet{Stan1999} 
undertook a more general study,
comparing the measurements of a set of roughly 20 
stars near SS Cyg
made by many visual observers
to the Johnson $V$ as a function of $(B-V)$.
He found a relationship 
\begin{equation}
 ({\rm visual - V}) = 0.21 * (B-V) \\
\end{equation}
which we plot with a dash-dotted line in
Figure
\ref{fig:comparevisual}.
The slope of this relationship
is consistent with that derived from the 
entire SN 2011fe dataset.

\section{Conclusion}

Our multicolor photometry suggests that SN 2011fe 
was a ``normal'' type Ia SN,
with a decline parameter
$\Delta m_{15}(B) = 1.21 \pm 0.03$ mag.
After correcting for extinction 
and adopting a distance modulus to M101
of $(m-M) = 29.10$ mag,
we find absolute magnitudes
of
$M_B = -19.21$,
$M_V = -19.19$,
$M_R = -19.18$
and 
$M_I = -18.94$,
which provide further evidence that this event
was ``normal'' in its optical properties.
As such, it should serve as an 
exemplar of the SNe which can act as
standard-izable candles for cosmological studies.
A comparison of the visual and CCD $V$-band measurements
of SN 2011fe
reveal systematic differences as a function of color
which are similar to those found for
other type Ia SNe and for stars in general.

\acknowledgements
We acknowledge with thanks the variable star observations 
from the AAVSO International Database contributed by observers 
worldwide and used in this research.
We thank Arne Henden and the staff at AAVSO for 
making special efforts to provide a sequence of comparison
stars near M101, and for helping to coordinate 
efforts to study this particular variable star.
MWR is grateful for the continued support of the
RIT Observatory by RIT and its College of Science.
Without the Palomar Transient Factory, the astronomical
community would not have received such early 
notice of this explosion.
We thank the anonymous referee for his comments.

{\it Facility:} \facility{AAVSO}

\newpage


\begin{thebibliography}{}
  \bibitem[Baade \& Zwicky\ (1938)] {Baad1938} 
           Baade, W. \& Zwicky, F.: 1938, ApJ, 88, 411
  \bibitem[Beyer\ (1939)] {Beye1939} 
           Beyer, M.: 1939, AN, 268, 341
  \bibitem[Cook, Aaronson \& Illingworth\ (1986)] {Cook1986} 
           Cook, K. H., Aaronson, M., \& Illingworth, G.: 1986, ApJ, 301, L45
  \bibitem[Hamuy et al.\ (1991)] {Hamu1991} 
           Hamuy, M., et al.: 1991, AJ, 102, 208
  \bibitem[Hamuy et al.\ (1996)] {Hamu1996} 
           Hamuy, M., et al.: 1996, AJ, 112, 2391
  \bibitem[Henden (2012)] {Hend2012} 
           Henden, A. A.: 2012, Observations from the 
                AAVSO International Database, private communication.
  \bibitem[Honeycutt (1992)] {Hone1992} 
           Honeycutt, R. K.: 1992, PASP, 104, 435
  \bibitem[Jacoby \& Pierce\ (1996)] {Jaco1996} 
           Jacoby, G. H. \& Pierce, M. J.: 1996, AJ, 112, 723
  \bibitem[Kirshner et al.\ (1973)] {Kirs1973} 
           Kirshner, R. P. et al.: 1973, 180, L97
  \bibitem[Landolt (1992)] {Land1992} 
           Landolt, A. U.: 1992, AJ, 104, 340
  \bibitem[Law et al. (2009)] {Law2009} 
           Law, N. M., et al.: 2009, PASP, 121, 1395
  \bibitem[Li et al. (2011)] {Li2011} 
           Li, W., et al.: 2011, Nature, 480, 348
  \bibitem[Lira\ (1995)] {Lira1995} 
           Lira, P.: 1995, Master's thesis, Univ. Chile
  \bibitem[Marion\ (2011)] {Mari2011} 
           Marion, H.: 2011, ATel, 3599, 1
  \bibitem[Munari \& Zwitter (1997)] {Muna1997} 
           Munari, U., \& Zwitter, T.: 1997, A\&A, 318, 269
  \bibitem[Nugent et al.\ (2011a)] {Nuge2011a} 
           Nugent, P. E., et al.: 2011a, ATel, 3581, 1
  \bibitem[Nugent et al.\ (2011b)] {Nuge2011b} 
           Nugent, P. E., et al.: 2011b, Nature, 480, 344
  \bibitem[Patat et al.\ (2011)] {Pata2011} 
           Patat, F., et al.: 2011, arXiv, 1112.0247
  \bibitem[Perlmutter et al.\ (1997)] {Perl1997} 
           Perlmutter, S., et al.: 1997, ApJ, 483, 565
  \bibitem[Pierce \& Jacoby\ (1995)] {Pier1995} 
           Pierce, M. J. \& Jacoby, G. H.: 1995, AJ, 110, 2885
  \bibitem[Phillips\ (1993)] {Phil1993} 
           Phillips, M. M.: 1993, ApJ, 413, L105
  \bibitem[Phillips et al.\ (1992)] {Phil1992} 
           Phillips, M. M., et al.: 1992, AJ, 103, 1632
  \bibitem[Phillips et al.\ (1999)] {Phil1999} 
           Phillips, M. M., et al.: 1999, AJ, 118, 1776
  \bibitem[Prieto, Rest \& Suntzeff\ (2006)] {Prie2006} 
           Prieto, J. L., Rest, A., \& Suntzeff, N. B.: 2006, ApJ, 647, 501
  \bibitem[Rau et al.\ (2009)] {Rau2009} 
           Rau, A., et al.: 2009, PASP, 121, 1334
  \bibitem[Riess, Press \& Kirshner\ (1996)] {Ries1996} 
           Riess, A. G., Press, W. H., \& Kirshner, R. P.: 1996, ApJ, 473, 88
  \bibitem[Richmond et al.\ (1995)] {Rich1995} 
           Richmond, M. W., et al.: 1995, AJ, 109, 2121
  \bibitem[Sakai et al.\ (2004)] {Saka2004} 
           Sakai, S., Ferrarese, L., Kennicutt, R. C., Saha, A.: 2004, ApJ, 608, 42
  \bibitem[Sandage \& Tammann (1974)] {Sand1974} 
           Sandage, A., \& Tammann, G. A.: 1974, ApJ, 194, 223
  \bibitem[Schlegel, Finkbeiner \& Davis (1998)] {Schl1998}
           Schlegel, D. J., Finkbeiner, D. P., \& Davis, M.: 
                   1998, ApJS, 500, 525
  \bibitem[Shappee \& Stanek\ (2011)] {Shap2011} 
           Shappee, B. J., \& Stanek, K. Z.: 2011, ApJ, 733, 124
  \bibitem[Simonson\ (2011)] {Simo2011} 
           Simonson, M.: 2011, BAAS, 218, 126.02
  \bibitem[Stanishev et al.\ (2007)] {Stan2007} 
           Stanishev, V., et al.: 2007, A\&A, 469, 645
  \bibitem[Stanton\ (1999)] {Stan1999} 
           Stanton, R. H.: 1999, JAVSO, 27, 97
  \bibitem[Treffers \& Richmond (1989)] {Tref1989} 
           Treffers, R. R., \& Richmond, M. W.: 1989, PASP, 101, 725
  \bibitem[Wells et al.\ (1994)] {Well1994} 
           Wells, L. A., et al.: 1994, AJ, 108, 2233

\end{thebibliography}
\end{document}